\documentclass[jkps,showpacs,showkeys,preprint]{revtex4}
\usepackage[dvips]{graphicx}
\usepackage{latexsym}
\begin{document}


\title{Entropy of 2+1 de Sitter space with the GUP}
 \author{Wontae \surname{Kim}}
   \email{wtkim@sogang.ac.kr}
 \affiliation{Department of Physics and Center for Quantum Spacetime,\\
  Sogang University, C.P.O. Box 1142, Seoul 100-611, Korea}
 \author{Yong-Wan \surname{Kim}}
   \email{ywkim@pcu.ac.kr}
 \affiliation{National Creative Research Initiative Center for Controlling
Optical Chaos, Pai-Chai University, Daejeon 302-735, Korea}
 \author{Young-Jai Park}
 \email{yjpark@sogang.ac.kr}
 \affiliation{Department of Physics and Center for Quantum Spacetime,\\
  Sogang University, Seoul 121-742, Korea}
\date[]{Received May 10 2006}

\begin{abstract}
By introducing the generalized uncertainty principle (GUP) on the
quantum state density, we calculate the statistical entropy of a
scalar field on the background of (2+1)-dimensional de Sitter space
without artificial cutoff. The desired entropy proportional to the
horizon perimeter is obtained.
\end{abstract}



\maketitle

\section{Introduction}
Three decades ago, Bekenstein suggested that the entropy of a black
hole is proportional to the area of the horizon through the
thermodynamic analogy \cite{bek}. Subsequently, Hawking showed that
the entropy of the Schwarzschild black hole satisfies exactly the
area law by means of Hawking radiation based on the quantum field
theory \cite{haw}. After their works, 't Hooft investigated the
statistical properties of a scalar field outside the horizon of a
Schwarzschild black hole by using the brick wall method (BWM) with
the Heisenberg uncertainty principle (HUP) \cite{tho}. The entropy
proportional to the horizon area is obtained, but a brick wall
cutoff, which was introduced to remove the divergence of the density
of states, looks unnatural. This method has been used to study the
statistical property of bosonic and fermionic fields in various
black holes \cite{gm,btz,kkps}, and it is found that the general
expression of the black hole entropy consists of the term, which is
proportional to the area of the horizon, and the divergent
logarithmic term. Although this BWM is useful for various models,
some difficulties may arise because it is assumed that there exists
a thermal equilibrium between the black hole and the external field
even in a large spatial region. Obviously, this method cannot be
applied to a non-equilibrium system such as a system of
non-stationary space-time with two horizons because the two horizons
have different temperatures and the thermodynamical laws are also
invalid there. Solving these problems, an improved brick-wall method
(IBWM) has been introduced by taking the thin-layer outside the
event horizon of a black hole as the integral region \cite{lz}. In
the thin-layer, local thermal equilibrium exists and the divergent
term due to large distance does not appear any more. However, the
IBWM does not still essentially solve the difficulties including the
artificial cutoffs.

On the other hand, many efforts have been devoted to the generalized
uncertainty relations \cite{gup}, and its consequences, especially
the effect on the density of states. Recently, in Refs.
\cite{li,liu}, the authors calculated the entropy of a black hole by
using the new equation of state density motivated by the generalized
uncertainty principle (GUP) \cite{gup}. As a result, the serious
divergence of the just vicinity near the horizon was removed.

However, most of statistical entropy calculations have been done for
the asymptotically flat cases by using the BWM and IBWM with the HUP
or by using the GUP. Up to now, the statistical entropy \cite{ms} of
the 2+1 de Sitter (DS) space, which has a cosmological horizon and
asymptotically non-flat spacetime, was only studied by using the BWM
\cite{wtk} and IBWM \cite{lo} with the HUP.

In this paper, we study the entropy of the black hole in the 2+1 DS
space. Firstly, we briefly recapitulate the previous results in BWM
and IBWM with the HUP. But, we will avoid the difficulty in solving
the Klein-Gordon wave equation by using the quantum statistical
method. Next, we derive the free energy of a massive scalar field on
the DS space background directly by using the new equation of state
density motivated by the GUP \cite{li,liu} in the quantum gravity.
Finally, we calculate the quantum entropy of the black hole via the
relation between free energy and entropy. As a result, we obtain the
desired entropy proportional to the horizon perimeter without any
artificial cutoff and any little mass approximation.

\section{Scalar field on 2+1 de Sitter Background}

Let us start with the following action
\begin{equation}
\label{action}
I=\frac{1}{2\pi} \int d^3 x \sqrt{-g} \left[R -\frac{2}{l^2} \right] ,
\end{equation}
where $\Lambda= \frac{1}{l^2}$ is a cosmological constant.
Then the classical equation of motion yields the DS metric as
\begin{eqnarray}
\label{metric}
ds^2 &=& -f(r) dt^2 +\frac{1}{f(r)} dr^2 + r^2 d \theta^2, \\
f(r) &=& \left(1- \frac{r^2}{l^2} \right).
\end{eqnarray}
The horizon is located at $r\equiv r_H=l$ and our spacetime is
bounded by the horizon as the two-dimensional cavity within the
inner space of the horizon ($0 \le r \le l$) in contrast to the BTZ
case \cite {btz} where the spacetime is defined within the outer
space of the horizon ($l \le r < \infty$). The inverse of Hawking
temperature is given by
\begin{equation}
\label{ht}
\beta_H = 2\pi l.
\end{equation}

In this DS background, let us first consider a scalar field with
mass $\mu$, which satisfies the Klein-Gordon equation
\begin{equation}
\label{kg} \frac{1}{\sqrt{-g}} \partial_{\mu}( \sqrt{-g} g^{\mu \nu}
\partial_{\nu} \Phi) - \mu^{2} \Phi = 0.
\end{equation}
Substituting the wave function $\Phi(r, \theta, t) = e^{-i\omega
t}\Psi(r, {\theta})$, we find that this Klein-Gordon equation
becomes
\begin{equation}
\label{rtheta0} \frac{\partial^{2} \Psi}{{\partial {r}}^2}  +
\left( \frac{1}{f} \frac{\partial f}{\partial r} +
\frac{1}{r}\right) \frac{\partial {\Psi}}{\partial {r}} +
\frac{1}{f} \left({\frac{1}{r^2}}\frac{{\partial}^{2}}{\partial
{\theta}^{2}} + \frac{\omega^{2}}{f} - \mu^{2} \right)\Psi = 0.
\end{equation}
By using the Wenzel-Kramers-Brillouin (WKB) approximation \cite{tho}
with $\Psi \sim exp[iS(r,\theta)]$, we have
\begin{equation}
\label{wkb} {p_{r}}^{2} = \frac{1}{f}\left(\frac{\omega^{2}}{f} -
\mu^{2} - \frac{p^2_\theta}{r^2}\right),
\end{equation}
where
\begin{equation}
\label{mom} p_{r} = \frac{\partial S}{\partial r},~
 p_{\theta} = \frac{\partial S}{\partial \theta}.
\end{equation}
On the other hand, we also obtain the square module momentum
\begin{equation}
\label{smom} p^{2} = p_{i}p^{i} = g^{rr}{p_{r}}^{2} + g^{\theta
\theta}{p_{\theta}}^{2} = \frac{\omega^{2}}{f} - \mu^{2}.
\end{equation}
Then, the area in the momentum phase space is given by
\begin{eqnarray}
A_{p}(r) &=& \int dp_r dp_\theta =
\pi\sqrt{\frac{1}{f}(\frac{\omega^2}{f}-\mu^2)}
\cdot\sqrt{r^2(\frac{\omega^2}{f}-\mu^2)} \\
&=& \pi \frac{ r}{\sqrt{f}} (\frac{\omega^2}{f}-\mu^2)
\end{eqnarray}
with the condition $\omega\geq\mu\sqrt{f}$.

\section{Brick Wall Models with HUP}
\subsection{Original Brick Wall Model}

According to the BWM with the HUP, let us briefly recapitulate the
previous work \cite{wtk} in the 2+1 de Sitter space. However, we
would like to avoid the difficulty of solving the wave equation by
using the quantum statistical method. The usual position-momentum
uncertainty relation followed by the HUP is given by
\begin{equation}
{\Delta x} {\Delta p} \ge \frac{\hbar}{2}.
\end{equation}
From now on we take the units $G=c=\hbar=k_{B}\equiv 1$. When
gravity is ignored, the number of quantum states in a volume element
in phase cell space based on the HUP in the 2+1 de Sitter space is
given by
\begin{equation}
\label{dn} dn = \left(\frac{dr dp_{r}}{2\pi}\right)
\left(\frac{d\theta dp_{\theta}}{2\pi}\right) = \frac{d^2 x d^2
p}{(2\pi)^2} ,
\end{equation}
where one quantum state corresponds to a cell of volume $(2\pi)^2$
in the phase space \cite{li,liu}. Then, the number of quantum states
with energy less than $\omega$ is given bya
\begin{eqnarray}
\label{nqs} n_{O}(\omega) &=& \frac{1}{(2\pi)^2}\int dr d\theta
dp_{r} dp_{\theta} = \frac{1}{(2\pi)^2}\int dr d\theta A_{p}(r)\nonumber   \\
&=& \frac{1}{2}\int^{r_{H}-\epsilon}_{L} dr \frac{r}{\sqrt{f}}
\left(\frac{{\omega}^2}{f}- \mu^{2}\right).
\end{eqnarray}
Note that $\epsilon$ and $L$ are ultraviolet and infrared
regulators, respectively, where $\epsilon > 0$ and $0 \leq L < r_H
-\epsilon$.

On the other hand, for the bosonic case the free energy at inverse
temperature $\beta$ is given by
\begin{equation}
\label{def} e^{-\beta F} = \prod_K
                \left[ 1 - e^{-\beta \omega_K} \right]^{-1}~,
\end{equation}
where $K$ represents the set of quantum numbers. By using Eq.
(\ref{nqs}), the free energy can be rewritten as
\begin{eqnarray}
\label{OfreeE0}
 F_{O} &=& \frac{1}{\beta}\sum_K \ln \left[ 1 - e^{-\beta \omega_K} \right]
   ~\approx ~\frac{1}{\beta} \int dn_{O}(\omega) ~\ln
            \left[ 1 - e^{-\beta \omega} \right]  \nonumber   \\
   &=& - \int^{\infty}_{\mu\sqrt{f}} d \omega \frac{n_{O}(\omega)}{e^{\beta\omega} -1} \nonumber  \\
   &=& - \frac{1}{2} \int^{r_{H}-\epsilon}_{L} dr \frac{r}{\sqrt{f}}
   \int^{\infty}_{\mu\sqrt{f}} d\omega
   \frac{\left(\frac{{\omega}^2}{f}- \mu^{2}\right)}{(e^{\beta\omega} -1)}.
\end{eqnarray}
Here we have taken the continuum limit in the first line and
integrated by parts in the second line.

Now, let us evaluate the entropy for the scalar field, which can be
obtained from the free energy (\ref{OfreeE0}) at the Hawking
temperature; then the entropy is
\begin{eqnarray}
\label{entropy01} S_{O} &=&  \beta^2 \frac{\partial F_O}{\partial
\beta} \nonumber \\
&=& \frac{\beta^{2}}{2} \int^{r_{H}-\epsilon}_{L}dr \frac{r}{
\sqrt{f}} \int^\infty_{\mu\sqrt{f}} d\omega  \frac{\omega e^{\beta
\omega} \left(\frac{{\omega}^2}{f}- \mu^{2}\right)}{(e^{\beta
\omega} -1)^{2}}.
\end{eqnarray}
Note that at this stage it is difficult to carry out the analytic
integral about $\omega$ because the value of ${\mu\sqrt{f}}$ varies
depending on $r$ in the wide range $(L, r_{H}-\epsilon)$.

For the case of the massless limit, the entropy becomes
\begin{eqnarray}
\label{entropy02} S_{O} &=& \frac{\beta^{-2}}{2}
\int^{r_{H}-\epsilon}_{L} \frac{r dr}{f
\sqrt{f}} \int^\infty_0 \frac{e^{x}x^{3}dx}{(e^{x} -1)^2} \nonumber \\
&=& \frac{1}{2}\pi a  \left(\frac{l}{\sqrt{l^2-(r_H -\epsilon)^2}}-
                \frac{l}{\sqrt{l^2-L^2}}  \right),
\end{eqnarray}
where $x=\beta \omega$ and the constant is defined by $a \equiv
\frac{3 \zeta(3)}{2 \pi^3}$. Note that this result is exactly the
same as that of the previous work \cite {lo}, which was obtained
through considering of the number of modes according to the
semiclassical quantization rule. Then, when $\epsilon \rightarrow
0$, the dominant contribution term to the entropy is given by
\begin{equation}
\label{Lentropy} S_{O} \approx \frac{\pi a}{2\sqrt{2}}
\sqrt{\frac{l}{\epsilon}}.
\end{equation}
As a result, the ultraviolet divergence of the entropy comes from
near horizon ($r_{H}-\epsilon \le r \le r_{H}$) as the BTZ case
($r_{H} \le r \le r_{H} + \epsilon$). Moreover, the invariant
distance of the brick wall from the horizon at $ r= r_H = l$ is
related to the ultraviolet cutoff as
\begin{equation}
\label{invariant0} \tilde{\epsilon}= \int_{r_H -\epsilon}^{r_H}
\frac{dr}{\sqrt{f(r)}} = l \left(\frac{\pi}{2}
-\sin^{-1}\frac{l-\epsilon}{l} \right).
\end{equation}
Then, the entropy (\ref{entropy02}), which is always positive, can
be represented in terms of the invariant cutoff (\ref{invariant0})
as follows
\begin{equation}
\label{entropy} S_{O} = \frac{1}{2} \pi a \left( \frac{1- \sin
\frac{\tilde{\epsilon}}{l}}{\sin \frac{\tilde{\epsilon}}{l}} \right)
\equiv \frac{1}{2} \pi a~ s(\tilde{\epsilon},l).
\end{equation}
On the other hand, the infrared cutoff can be simply fixed as $L=0$
without loss of generality because there does not exist any infrared
divergence in the DS space where the spacetime is bounded within the
inner space of the horizon ($0 \le r \le r_{H}$) in contrast to the
BTZ black hole case where the spacetime is defined within the outer
space of the horizon ($r_{H} \le r < \infty$). Furthermore, if we
choose the cutoff $\tilde{\epsilon}$ as $a \equiv l /
s(\tilde{\epsilon},l)$, then the entropy can be rewritten by the
perimeter law $S_{O} = (2 \pi r_H)/4$. Note that for $l \gg
\tilde{\epsilon}$, the invariant cutoff is simply written as
$\tilde{\epsilon} \approx a$ that does not depend explicitly on
$r_{H}=l$.

\subsection{Improved Brick Wall Model}
Although the BMW with the HUP has contributed a great deal to the
understanding and calculating of the entropy of a black hole, there
are generally some drawbacks in it, such as little mass
approximation, neglecting logarithm term and artificial cutoffs.
Moreover, the fundamental problem is why the entropy of fields
surrounding the black hole is the entropy of the black hole itself
since the event horizon is the characteristic of a black hole.
Therefore, the entropy calculating of a black hole should be only
related to its horizon. Due to this reason and the fact that the
density of quantum states near the horizon is divergent, the BWM
have been improved to take only the entropy of a thin-layer near the
event horizon of a black hole avoiding the drawbacks in the original
BMW including the little mass approximation.

Now, according to the IBWM \cite{lz,lo}, let us summarize the
improved results in the 2+1 DS space comparing with those of the
original BWM. By just replacing the integral range $(L,
r_H-\epsilon)$ of the BWM in the entropy (\ref{entropy01}) with
$(r_H-\epsilon_{2}, r_H-\epsilon_{1})$, we have the entropy for a
massive scalar field as follows
\begin{eqnarray}
\label{Ientropy01} S_{T} &=& \frac{\beta^2}{2}\int^{r_H-
\epsilon_{1}}_{r_H- \epsilon_{2}}dr \frac{r}{\sqrt{f}}
\int^{\infty}_{\mu\sqrt{f}} d\omega \frac{\omega
e^{\beta\omega}(\frac{\omega^2}{f}-\mu^2)}
{(e^{\beta \omega}-1)^2} \nonumber \\
&\equiv&
\frac{\beta^2}{2}\int^{r_H-\epsilon_{1}}_{r_H-\epsilon_{2}}dr
\frac{r}{\sqrt{f}}~\Lambda_T,
\end{eqnarray}
where $\epsilon_{i}(i=1,2)$ with $\epsilon_{1} < \epsilon_{2}$
represent the coordinate distances from the horizon to the nearest
and more distant boundary, respectively, of the thin-layer. Since
$f\rightarrow 0$ in the near horizon range of $(r_H-\epsilon_{2},
r_H-\epsilon_{1})$, without any little mass approximation, the
integral about $\omega$ is reduced to
\begin{eqnarray}
\Lambda_T =\int^{\infty}_{0} d\omega
\frac{f^{-1}\omega^3}{(1-e^{-\beta\omega})(e^{\beta\omega}-1)}=\int^{\infty}_{0}
dx \frac{f^{-1}\beta^{-4}x^3}{(1-e^{-x})(e^x-1)},
\end{eqnarray}
where $x=\beta\omega$. Then, the integration gives explicitly the
result as
\begin{equation}
\label{Ientropy02} S_{T} = \frac{1}{2}\pi a
\left(\frac{l}{\sqrt{l^2-(r_H -\epsilon_{1})^2}}-
                \frac{l}{\sqrt{l^2-(r_{H}-\epsilon_{2})^2}}  \right).
\end{equation}
This result shows that the entropy behaves as
$1/\sqrt{\epsilon_{i}}$ at $\epsilon_{i} \rightarrow 0$, which
correspond to the ultraviolet divergences of the entropy.

On the other hand, the invariant distances of the thin-layer from
the horizon at $ r= r_H = l$ are related to the ultraviolet cutoffs
$\epsilon_{i}$ as
\begin{equation}
\label{Iinvariant0} \tilde{\epsilon_{i}}=\int_{{r_H}
-\epsilon_{i}}^{r_H} \frac{dr}{\sqrt{f(r)}}= l \left(\frac{\pi}{2}
-\sin^{-1}\frac{l-\epsilon_{i}}{l} \right).
\end{equation}

Then, the entropy (\ref{Ientropy02}) can be represented in terms of
the invariant cutoffs (\ref{Iinvariant0}) as follows
\begin{equation}
\label{Ientropy03} S_{T} = \frac{1}{2} \pi a \left( \frac{\sin
\frac{\tilde{\epsilon_{2}}}{l}- \sin\frac{\tilde{\epsilon_{1}}}{l}}
{\sin\frac{\tilde{\epsilon_{1}}}{l}\sin
\frac{\tilde{\epsilon_{2}}}{l}} \right) \equiv \frac{1}{2} \pi a
~s(\tilde{\epsilon_{i}},l).
\end{equation}
Note that there does not exist any infrared divergence even though
we consider a massive scalar field, and the entropy $S_{T}$ is
always positive since $\tilde{\epsilon_{1}} < \tilde{\epsilon_{2}}$.
Furthermore, if we choose the cutoffs as $a = l/
s(\tilde{\epsilon_{i}},l)$, then the entropy can be also rewritten
by the perimeter law $S_{T} = (2 \pi r_H)/4$ as the BWM case.

It seems to be appropriate to comment on the entropy relation
between the BWM and IBWM. For $l \gg \tilde{\epsilon_{i}}$, we could
choose the cutoffs in the IBWM  as $\tilde{\epsilon_{2}} \equiv
2\tilde{\epsilon}$ and $\tilde{\epsilon_{1}} \equiv
\tilde{\epsilon}$ without loss of generality. Then, the value $a$
satisfying the perimeter law $S_{T}$ becomes $2\tilde{\epsilon}$,
and this value does not also depend explicitly on $r_{H}=l$ as the
original BWM case. Furthermore,  the entropy $S_{O}$ with $a =
2\tilde{\epsilon}$ in Eq. (\ref{entropy}) becomes $S_{O} \approx 2
S_{T}$. This means that the entropy contribution of the whole rest
range $(0, r_{H}-2\epsilon)$ is equal to that of the near horizon
range $(r_{H}-2\epsilon, r_{H}-\epsilon)$.

\section{Entropy with Generalized Uncertainty Principle}

Recently, many efforts have been devoted to the generalized
uncertainty relation \cite{gup} given by
\begin{equation} {\Delta x} {\Delta p} \ge \frac{1}{2}\left(1 +
{\lambda}({\Delta p})^{2}\right).
\end{equation}
Then, since one can easily get ${\Delta x} \geq \sqrt{\lambda}$,
which gives the minimal length, it can be defined to be the
thickness of the thin-layer near horizon, which naturally plays a
role of the brick wall cutoff. Furthermore, based on the generalized
uncertainty relation, the volume of a phase cell in the de Sitter
space is changed from $(2{\pi})^{2}$ into
\begin{equation}
\label{gup} (2{\pi})^{2}(1 + {\lambda}{p^{2}})^{2},
\end{equation}
where $p^2 = p^{i}p_{i}~(i = r, \theta).$

From Eq. (\ref{gup}), the number of quantum states with energy less
than $\omega$ is given by
\begin{eqnarray}
\label{Tnqs} n_{I}(\omega) &=& \frac{1}{(2\pi)^2} \int dr d\theta
dp_{r} dp_{\theta} \frac{1}{\left(1+
{\lambda}(\frac{{\omega}^2}{f}- \mu^{2})\right)^2} \nonumber   \\
&=&\frac{1}{(2\pi)^2} \int dr d\theta \frac{1}{\left(1+
{\lambda}\left(\frac{{\omega}^2}{f}- \mu^{2}\right) \right)^2} A_{p}(r) \nonumber   \\
&=& \frac{1}{2}\int dr \frac{r}{\sqrt{f}}
\frac{\left(\frac{{\omega}^2}{f}- \mu^{2}\right)}{\left(1+
{\lambda}(\frac{{\omega}^2}{f}- \mu^{2})\right)^2}.
\end{eqnarray}
Note that it is convergent at the horizon without any artificial
cutoff due to the existence of the suppressing $\lambda$ term in the
denominator induced from the GUP. Then, by using Eq. (\ref{Tnqs}),
the free energy can be obtained as
\begin{eqnarray}
\label{TfreeE}
 F_{I} &=& - \int^{\infty}_{\mu\sqrt{f}} d\omega \frac{n_{I}(\omega)}{e^{\beta\omega} -1} \nonumber  \\
   &=& - \frac{1}{2} \int dr \frac{r}{\sqrt{f}}
   \int^{\infty}_{\mu\sqrt{f}} d\omega \frac{\left(\frac{{\omega}^2}{f}
   - \mu^{2}\right)}{(e^{\beta \omega} -1)\left(1+ \lambda
   (\frac{{\omega}^2}{f}-\mu^2)\right)^2}.
\end{eqnarray}
From this free energy, the entropy for the massive scalar field is
given by
\begin{eqnarray}
\label{Tentropy0} S_{I} &=& \beta^2 \frac{\partial F_I}{\partial
\beta} \nonumber \\
&=& \frac{\beta^2}{2} \int dr \frac{r}{\sqrt{f}}
\int^\infty_{\mu\sqrt{f}} d\omega \frac{\omega e^{\beta \omega}
\left(\frac{\omega^2}{f}- \mu^2\right)}
{(e^\omega-1)^2 (1+ \lambda (\frac{\omega^2}{f}- \mu^{2}))^2} \nonumber \\
&\equiv& \frac{\beta^2}{2} \int dr \frac{r}{\sqrt{f}} \Lambda_{I}.
\end{eqnarray}
Since $f \rightarrow 0$ near the event horizon, {\it i.e.}, in the
range of $(r_H-\epsilon, r_H)$, then, without any little mass
approximation, the integral about $\omega$ is reduced to

\begin{equation}
\label{Tentropy1} \Lambda_{I} = \int^\infty_0 dx \frac{f^{-1}
\beta^{-4} x^{3}}{(1-e^{-x})(e^{x}
-1)\left(1+\frac{\lambda}{\beta^{2}f}x^{2}\right)^2}, \nonumber \\
\end{equation}
where $x=\beta \omega$.

On the other hand, we are only interested in the contribution from
the just vicinity near the horizon, $(r_H - \epsilon, r_H)$, which
corresponds to a proper distance of order of the minimal length,
$\sqrt{\lambda}$. This is because the entropy closes to the upper
bound only in this vicinity, which it is just the vicinity neglected
by BWM and IBWM. We have
\begin{eqnarray}
\label{invariant} \sqrt{\lambda}=\int_{r_H -\epsilon}^{r_H}
                          \frac{dr}{\sqrt{f(r)}}
                = \int_{r_H -\epsilon}^{r_H}
                          \frac{dr}{\sqrt{2\kappa(r_{H}-r)}}
                = \sqrt{\frac{2\epsilon}{\kappa}},
\end{eqnarray}
where $\kappa$ is the surface gravity at the horizon of black hole
and it is identified as $\kappa = 2\pi \beta$.

Now, let us rewrite Eq. (\ref{Tentropy0}) as
\begin{equation}
\label{Sdef} S_{I} = \frac{1}{2\lambda}
\int^{r_H}_{r_{H}-\epsilon} dr \frac{r}{\sqrt{f}} \Lambda_{I},
\end{equation}
where
\begin{equation}
\label{Gdef} \Lambda_{I} = \int^\infty_0 dX
\frac{b^{2}X^{3}}{(e^{\frac{b}{2} X}-e^{-
\frac{b}{2}X})^{2}(1+X^2)^2}
\end{equation}
with $x=\beta\sqrt{\frac{f}{\lambda}}X \equiv bX$. Since
$r\rightarrow r_{H}$, $b \rightarrow 0$, removable pole becomes
$(e^{\frac{b}{2}X}-e^{- \frac{b}{2}X})^{2}\approx b^{2}X^{2} +
O(b^{3})$. Then, the integral equation (\ref{Gdef}) can be easily
solved without the help of the complex residue theorem as follows
\begin{equation}
\label{GI} \Lambda_{I} \cong \int^\infty_0 \frac{X dX}{(1+X^2)^2} =
\frac{1}{2}.
\end{equation}
Finally, when $r\rightarrow r_{H}$, we get the entropy as follows
\begin{equation}
\label{finalS} S_{I} = \frac{1}{2\lambda}\cdot r_H\sqrt{\lambda}
\cdot \frac{1}{2} = \frac{2\pi r_{H}}{8\pi \sqrt{\lambda}}.
\end{equation}
Note that there is no divergence within the just vicinity near the
horizon due to the effect of the generalized uncertainty relation on
the quantum states. Furthermore, if we assume $\lambda = \alpha
{l_{P}}^{2}$, where $l_{P}$ is Planck length, and in the system of
Planck units $l_{P}= 1$, then the entropy can be rewritten by the
desired perimeter law $S_{I} = \frac{1}{4} (2\pi r_{H})$ with
$\alpha =\frac{1}{4\pi^2}$.

It seems to be appropriate to comment on the entropy
(\ref{Lentropy}) of the original BWM comparing with the entropy
(\ref{finalS}), which is effectively considered the contribution
inside of the brick wall. From the Eq. (\ref{invariant0}), if we
take the brick wall cutoff $\tilde{\epsilon}$ as the minimal length
$\sqrt{\lambda}$ induced by the GUP, we have the relation $\epsilon
= \frac{\lambda}{2l}$. Then, we effectively obtain the entropy
contribution outside of the brick wall as follows
\begin{equation}
\label{BS0} S_{O} \approx \frac{3 \zeta(3)}{\pi^{2}} \frac{2\pi
r_{H}}{8\pi \sqrt{\lambda}} =  \frac{3 \zeta(3)}{\pi^{2}} S_{I}.
\end{equation}

In summary, we have investigated the massive scalar field within the
just vicinity near the horizon of a static black hole in the 2+1 de
Sitter space by using the generalized uncertainty principle. In
contrast to the cases of the BWM and IBWM, we have obtained the
desired entropy proportional to the horizon perimeter without any
artificial cutoff and any little mass approximation, simultaneously.

\begin{acknowledgments}
This work is supported by the Science Research Center Program of the
Korea Science and Engineering Foundation through the Center for
Quantum Spacetime of Sogang University with grant number
R11-2005-021. Y.-J. Park is supported by the Korea Research
Foundation Grant funded by the Korean Government
(KRF-2005-015-C00105).
\end{acknowledgments}

\end{document}